# Topological Dark Spots of Electric Near Field in Metal Structures

Tong Fu[1], Qing Tong[1], Shiqi Jia[1], and Shubo Wang[1,*]

[1]Department of Physics, City University of Hong Kong, Tat Chee Avenue, Kowloon, Hong Kong, China.

*Corresponding author. Email: shubwang@cityu.edu.hk

**ABSTRACT**: Electric dark spots are point singularities at which the electric field amplitude vanishes. These singularities usually emerge in real space accidentally and are unstable due to the vectorial property of the electric field. In this paper, we show that topologically protected electric dark spots can emerge in metal scatterers under external excitation. The material property of metal imposes a boundary condition that reduces the vectorial electric field on the metal surface to a scalar field. The phase singularity of this scalar field has zero amplitude and carries a well-defined topological charge corresponding to an electric dark spot. The topological electric dark spots give rise to the superoscillation phenomenon with a divergent local wavenumber. We uncover the global charge conservation property of the dark spots on the scatterers' surfaces and demonstrate their stability under different perturbations. We also demonstrate the manipulation of the dark spots' topological charge and spatial location. The results open a new avenue for nanophotonic near-field manipulations and may find applications in optical metrology, optical sensing, and super-resolution imaging.

## 1. INTRODUCTION

An electric dark spot is a singular point at which all components of the electric field vanish simultaneously[1]. One well-known example is the center of an optical vortex beam with zero intensity[2]. The electric dark spots carry fascinating properties, such as the circulation of the Poynting vector[1] and superoscillation[3-7]. They play an essential role in numerous applications, including objects hiding[8], super-resolution imaging[9-11], atoms trapping[12], etc. Thus, the generation and manipulation of electric dark spots have attracted growing interest in recent years[13-18].

In singular optics, electric dark spots are known as the V points, which is a type of high-order polarization singularities[2,19]. The generic polarization singularities are the C points at which the field is circularly polarized with an ill-defined major axis of the polarization ellipse[20]. These polarization singularities can emerge in both the near and far fields of optical systems[21-25]. In contrast to the C points, the V points are generally unstable and can easily split into multiple C points under perturbations. Thus, dark spots are usually generated in optical systems with spatial symmetries (e.g., mirror symmetry)[24]. Furthermore, dark lines (i.e., lines of dark spots) can only be observed in some highly symmetrical systems[8,26,27]. For example, rotational and translational symmetries can protect the electric dark lines in vortex beams[2,27,28]. Mirror and time-reversal symmetries can give rise to electric dark lines in photonic crystal slabs[8].

In the absence of symmetry protection, electric dark spots are only stable in higher dimensional space[1,29]. For instance, the electric dark spots of paraxial beams can stably exist in a synthetic four-dimensional space comprising three spatial dimensions and one parameter (e.g., wavelength) dimension[29]. In this case, a perturbation applied to the system simply shifts the dark spots along the parameter dimension. For a general three-component electric field, the dark spots can only stably reside in six-dimensional space[1]. This entails three more parameters besides the three spatial coordinates to guarantee the robustness of the dark spots. The availability and free tuning of these parameters pose significant challenges to realizing electric dark spots in general optical systems. Therefore, exploring simple and efficient approaches to generate stable electric dark spots are highly desirable and useful.

In this work, we demonstrate that stable dark spots of the electric near field can be induced in metal scatterers without symmetry, inspired by the concept of scalar topological photonics[30]. The boundary condition imposed by metals reduces the electric near field to a scalar field that ensures the topological nature of the dark spots. The dark spots are closely related to the emergence of near-field vortices and superoscillation. We determine the topological charges of the dark spots and reveal their global conservation on the scatterers' surfaces. We also demonstrate the stability of the dark spots under perturbations and the manipulation of their topological charges and spatial positions.

## 2. RESULTS AND DISCUSSION

**Topological dark spots of electric near field**

A general non-paraxial electric field in three-dimensional (3D) real space can be written as $\mathbf{E}(\mathbf{r}) = \mathbf{E}_\Re(\mathbf{r}) + i\mathbf{E}_\Im(\mathbf{r})$, where $\mathbf{E}_\Re(\mathbf{r}) = E_{\Re x}\hat{\mathbf{x}} + E_{\Re y}\hat{\mathbf{y}} + E_{\Re z}\hat{\mathbf{z}}$ and $\mathbf{E}_\Im(\mathbf{r}) = E_{\Im x}\hat{\mathbf{x}} + E_{\Im y}\hat{\mathbf{y}} + E_{\Im z}\hat{\mathbf{z}}$ are the real and the imaginary parts of the electric field vector, respectively. The emergence of the electric dark spots is a codimension-six phenomenon requiring $E_{\Re x} = E_{\Re y} = E_{\Re z} = E_{\Im x} = E_{\Im y} = E_{\Im z} = 0$. Yet, the real space is 3D, indicating that electric dark spots cannot stably exist in real space. For a paraxial beam propagating along the $z$ direction, we have $\mathbf{E}_\Re = E_{\Re x}\hat{\mathbf{x}} + E_{\Re y}\hat{\mathbf{y}}$ and $\mathbf{E}_\Im = E_{\Im x}\hat{\mathbf{x}} + E_{\Im y}\hat{\mathbf{y}}$. The corresponding electric dark spots are a codimension-four phenomenon, which still cannot survive in 3D real space under general perturbations. Undoubtedly, the only possible way to realize stable electric dark spots in 3D real space is to further reduce the codimension to two. This corresponds to a scalar electric field $E = E_\Re + iE_\Im$ whose zeros form one-dimensional lines in 3D real space (note that the dimension of the zeros is the difference between the dimension of the space and the codimension[31,32]). On a two-dimensional (2D) surface embedded in 3D real space, the zeros of the scalar electric field correspond to zero-dimensional dark spots[33]. This guides us to consider optical systems supporting scalar electric fields living on 2D surfaces, such as the metal structures with near perfect-electric-conductor (PEC) property. Under the PEC boundary condition, the electric near field is perpendicular to the metal surface and can be regarded as a scalar field defined on a 2D surface. Using full-wave numerical simulations with COMSOL Multiphysics, we will show that such systems can give rise to stable electric dark spots with a topological nature.

We first consider a PEC sphere with a radius of 100 nm under the illumination of a monochromatic plane wave $\mathbf{E}_0 = (\hat{\mathbf{x}} + i0.6\hat{\mathbf{y}})e^{ikz}$ (the time dependence factor $\exp(-i\omega t)$ is neglected hereafter). We choose elliptical polarization for the incident plane wave to ensure the asymmetry of the system. The frequency throughout the paper is set to be $f = 150$ THz unless otherwise specified. Under the excitation of the plane wave, the sphere can induce C lines (i.e., lines of C points) in the total electric near field $\mathbf{E}$, as shown by the two dark cyan lines in Figure 1a. The two C lines merge on the sphere surface and give rise to two electric dark spots at the north and south poles of the sphere, as marked by the solid and dashed squares in Figure 1a, respectively. The relative electric field strength $|\mathbf{E}|^2/|\mathbf{E}_0|^2$ is shown in Figure 1c. We notice that the two dark spots have a negligible electric field with $|\mathbf{E}|^2/|\mathbf{E}_0|^2 < 10^{-5}$. The emergence of the dark spots can be understood as follows. Under the boundary condition at the sphere surface, the electric field is linearly polarized with only a perpendicular component, i.e., the electric field is locally perpendicular to the surface. However, the electric field on the C lines is circularly polarized with two orthogonal field components. Consequently, the local electric field must vanish when the C lines connect to the sphere surface.

The electric dark spots in Figure 1 carry topological properties characterized by topological charges. The complex electric field can be expressed as $\mathbf{E}(\mathbf{r}) = (\mathbf{A}+i\mathbf{B})e^{i\arg(\psi)/2}$, where $\mathbf{A}$ and $\mathbf{B}$ are the major and minor axes of the polarization ellipse, and $\psi = \mathbf{E}\cdot\mathbf{E}$. For each C line in Figure 1a, the topological charge can be determined as

$$q = \frac{1}{2\pi}\oint \nabla\arg(\Psi) \cdot d\mathbf{r}, \qquad (1)$$

where $q$ is an integer, and the integral is carried out over a small loop enclosing the C line. This topological charge (also called phase index[24] since it is defined with the phase of $\psi$) is invariant when the small loop travels along the C line. Figure 1a shows the phase $\arg(\Psi)$ on two planes cutting the two C lines. We notice that the two C lines carry the same topological charges $q = \pm 1$ (the sign of the charge depends on the positive direction chosen for the integral in eq 1). The merging of the two identical charges gives rise to the electric dark spot of charge $q = +2$ ($q = -2$) on the north (south) pole of the sphere, assuming the direction of the integral in eq 1 is along the outward unit normal vector $\hat{\mathbf{n}}$ of the sphere surface. This is confirmed by the numerically simulated phase $\arg(\Psi)$ near the north and south dark spots, as shown in the upper and lower panels of Figure 1d, respectively

The electric dark spots are intimately related to topological near-field vortices and superoscillation. Optical superoscillation refers to the phenomenon that the magnitude of local wavevector $|\mathbf{k}|$ of optical fields is larger than the free space wavevector $|\mathbf{k}_0| = 2\pi/\lambda$, which has essential applications in super-resolution focusing and imaging[7,11,34]. Figure 1d shows that the local wavevector $\mathbf{k} = -iE^*(\nabla)E/|E|^2$ (denoted by the black arrows) near the two dark spots of the sphere, which circulates around the dark spots with the direction directly decided by the topological charge *q*. Therefore, the two dark spots carry opposite orbital angular momentum with respect to the direction of the surface normal $\hat{\mathbf{n}}$. This near field orbital angular momentum has potential application in the discrimination of chiral molecules[35,36]. Figure 1e shows the corresponding normalized wavenumber $|\mathbf{k}|/|\mathbf{k}_0|$, which diverges at the position of the dark spots and confirms the superoscillation property. Therefore, the wavenumber $|\mathbf{k}|$ can be evaluated as a figure of merit for the electric dark spots[18].

Unlike the electric dark spots in conventional systems, the dark spots here are protected by topology and are stable against small perturbations. This can be understood as follows. The electric field on the PEC surface is a complex scalar field $E(\mathbf{r}) = E_\Re(\mathbf{r}) + iE_\Im(\mathbf{r}) = |E(\mathbf{r})|e^{i\arg(\psi)/2}$, where $E_\Re$ and $E_\Im$ are the real and imaginary parts, respectively. The dark spots in Figure 1a correspond to the intersection points of the two lines $E_\Re = 0$ and $E_\Im = 0$[37], as shown in Figure 1b. Small perturbations can only move the intersection points on the PEC surface but cannot destroy them. The only way to eliminate the dark spots is to bring them together by some strong perturbations and make them annihilate.

The above interpretation based on $E_\Re$ and $E_\Im$ also provides an intuitive explanation for net topological charges of the dark spots. For the scalar electric field $E(\mathbf{r})$, eq 1 can be reduced to $q = 1/\pi \oint (E_\Re dE_\Im - E_\Im dE_\Re)/|E|^2 \cdot d\mathbf{r}$, which corresponds to two times of the winding number defined for a small loop enclosing the dark spot, as shown in the upper panels of Figure 1f. Since the lines of $E_\Re = 0$ and $E_\Im = 0$ are closed loops on the surface of the sphere, their intersection points must emerge or be annihilated in pairs with opposite charges (i.e., opposite winding numbers). Therefore, the net topological charges of the surface dark spots always satisfy

$$\sum_{i=1}^{n} q_i = 0, \qquad (2)$$

where *n* is the number of the electric dark spots and $q_i$ is the topological charge of the *i*th dark spot. Figure 1f shows four examples with *n* = 0 (type i), *n* = 2 (type ii), *n* = 4 (type iii), and *n* = 6 (type iv). In all the examples, the net charge is zero. The case of Figure 1b belongs to the type ii.

**Robustness of the dark spots upon perturbations**

To demonstrate the robustness of the electric dark spots, we consider the PEC scatterers with different geometries, as shown in Figure 2. The incident wave is the same as that in Figure 1. The numerically determined electric dark spots and the lines of $E_\Re = 0$ and $E_\Im = 0$ are shown on the scatterers' surface. Figure 2a shows that there are two dark spots emerging on the surface of the nanorod, similar to the case of the sphere in Figure 1. In addition, when the nanorod (Figure 2a) undergoes bending (Figure 2b) and then twisting (Figure 2c), the two dark spots remain stable and the topological configuration is always of type ii in Figure 1f, although their positions are different. Figure 2d shows the dark spots emerging on two PEC spheres that are placed close to each other. Despite the near-field coupling between the two spheres, dark spots generated on each sphere are similar to the single-sphere case in

Figure 1. The configurations of the dark spots are similar when the relative position of the spheres is changed, as shown in Figure 2e. It is well known that a metal dimer can induce electric hot spot in the nanogap[38]. Our results show that metal dimmers can also generate stable electric dark spots on their surfaces. We also verify the robustness of the dark spots in the presence of a substrate, as shown in Figure 2f, where the PEC sphere sits on a $SiO_2$ substrate of infinitely large thickness. Numerical results show that the dark spots and their topological configuration remain unaffected by the substrate. These results demonstrate that the electric dark spots on the PEC surface are indeed topologically protected and robust against small perturbations.

**Topological transition of the dark spots**

While the electric dark spots are stable against small perturbations, they can disappear when strong perturbations are applied to induce topological transitions. Such a scenario can happen when we deliberately tune the geometry parameters of the structures. Figure 3a shows the nanorod bent into a circle arc shape (similar to the so-called "split ring") with the opening defined by the angle $\alpha$. From left to right, we set $\alpha = 0.9\pi$, $\alpha = 0.66\pi$, and $\alpha = 0.6\pi$, respectively. The numerically simulated electric dark spots and the lines of $E_\Re = 0$ and $E_\Im = 0$ are shown on the surface. We see that the configuration of the dark spots changes from type ii to type i when the angle $\alpha$ reduces from $0.9\pi$ to $0.6\pi$. Meanwhile, the two dark spots of opposite charges gradually approach each other and then disappear, corresponding to a topological transition. Figure 3b shows the distribution of $|\mathbf{E}|^2/|\mathbf{E}_0|^2$ in the three cases, the minimum value increases from $\sim 10^{-5}$ to $\sim 4\times 10^{-3}$ after the topological transition. The topological transition can be clearly understood based on the C line configuration and the phase $\arg(\Psi)$ of the electric field. As shown in Figure 3c, the two dark spots correspond to two phase singularities at which the C lines merge. As $\alpha$ reduces, the two singularities carrying opposite charges approach each other and annihilate. Accordingly, the three C lines become one continuous C line without touching the PEC surface (Figure 3c right panel). Furthermore, the magnitude of the local wavevector $|\mathbf{k}|/|\mathbf{k}_0|$ changes from an infinite value to a finite value of $\sim 285$, as shown in Figure 3d. In the case of $\alpha = 0.6\pi$ (Figure 3d right panel), the lines of $E_\Re = 0$ and $E_\Im = 0$ are close to each other without crossings, and the local minimum of the electric field is still negligible compared to the incident field ($|\mathbf{E}|^2/|\mathbf{E}_0|^2 \sim 4\times 10^{-3}$). Thus, the point can be regarded as a quasi-dark spot with a large yet finite value of wavenumber.

**Higher-order dark spots**

Higher-order electric dark spots can appear in the systems with symmetry. We consider the PEC sphere excited by a circularly polarized Bessel beam propagating along the $z$-axis: $\mathbf{E}_0(\mathbf{r}) = \omega\mu_0 k_z \big[ i(\hat{\mathbf{x}} + i\hat{\mathbf{y}})J_m(k_t\rho)e^{im\beta} - \hat{\mathbf{z}}J_{m+1}(k_t\rho)e^{i(m+1)\beta} \big] e^{ik_z z}$, where $(\rho, \beta, z)$ are the cylindrical coordinates; $\mu_0$, $\omega$, $k_t$, $k_z$, denote the vacuum permeability, angular frequency, transverse and longitudinal wavenumbers, respectively; $J_m(k_t\rho)$ is the $m$-th order Bessel function of the first kind. The system has a rotational symmetry with respect to the $z$ axis, which ensures that the electric field satisfies[24,39]

$$R(\phi)\mathbf{E}(R(-\phi)\mathbf{r}) = e^{-i(m+1)\phi}\mathbf{E}(\mathbf{r}). \qquad (3)$$

Here, $R(\phi)$ denotes the rotation matrix with rotation angle $\phi$. On the $z$-axis, the above equation reduces to $R(\phi)\mathbf{E}(z\hat{\mathbf{z}}) = e^{-i(m+1)\phi}\mathbf{E}(z\hat{\mathbf{z}})$, indicating that $\mathbf{E}(z\hat{\mathbf{z}})$ must be either null or an eigenvector of the rotation matrix $R(\phi)$ with eigenvalue $e^{-i(m+1)\phi}$. It is known that the general rotation matrix $R(\phi)$ has three eigenvalues $e^{-i\phi}$, $e^{i\phi}$, and 1 with the corresponding eigenvectors $\mathbf{R} = \hat{\mathbf{x}} + i\hat{\mathbf{y}}$, $\mathbf{L} = \hat{\mathbf{x}} - i\hat{\mathbf{y}}$, and $\hat{\mathbf{z}}$. Therefore, we obtain the following conclusions: (1) When $m = 0$ or $-2$, the electric field on the $z$-axis is circularly polarized, corresponding to a higher-order C line, and electric dark spots with charge $q = \pm 2(m+1) = \pm 2$ must appear at the south and north poles of the sphere due to the PEC boundary condition; (2) When $m = -1$, the electric field on the $z$-axis is linearly polarized along the $z$ direction, corresponding to an L line (i.e., line of linear polarization), and no electric dark spot will emerge on the sphere's surface; (3) When $m > 0$ or $m < -2$, the electric field on $z$-axis must vanish, corresponding to a V line, and electric dark spots with charge $q = \pm 2(m+1)$ must appear at the south and north poles of the sphere.

To verify the above analysis, we conduct numerical simulations for the sphere excited by the Bessel beam with $m = +1$, 0, and $-1$. The results are summarized in Figure 4a-f. When $m = +1$, there are two electric dark spots with charge $q = \pm 4$ appear at the south and north poles, as shown in Figure 4a. Meanwhile, a V line appears on the central $z$-axis, as shown in Figure 4b. When $m = 0$, there are two electric dark spots with charge $q = \pm 2$ appearing at the south and north poles, as shown in Figure 4c. And, a C line appears on the $z$-axis, as shown in Figure 4d. When $m = -1$, no electric dark spot can be induced on the sphere's surface, as shown in Figure 4e, and an L line appears on the $z$-axis (Figure 4f). We note that the topological charge of the electric dark spots is always two times the total angular momentum quantum number of the incident Bessel beam up to a sign. There are no electric dark spots when the incident Bessel beam carries zero total angular momentum.

**Manipulating the dark spots' position**

The positions of electric dark spots on the PEC surface can be manipulated by controlling the properties of the incident plane wave. Here, we demonstrate numerically the manipulation of the dark spots' position. As shown in Figure 5a, we consider the electric dark spots generated on the surface of a nanorod under the excitation by a circularly polarized plane wave. The incident direction of the plane wave is described by the polar ($\theta$) and azimuthal ($\varphi$) angles. Figure 5b shows the simulated dark spots under different incident angles. As seen, the two dark spots move along different trajectories on the surface when the incident direction changes, and the topological charge of each dark spot remains unchanged during this process. The trajectories of the two dark spots are shown in Figure 5a as the blue and red lines. We notice that both trajectories have a helical shape but travel in opposite directions. The two dark spots swap their locations after the manipulation. We note that for the PEC sphere excited by a circularly polarized plane wave, the dark spots are located on a symmetric axis parallel to the incident direction, as in the case of Figure 4c. In principle, more complex trajectories of the dark spots can be achieved by controlling the properties of the incident light.

**Effect of material dispersion and loss**

In the above discussions, we assume the structures are made of a PEC. The physics equally applies to real metals (e.g., copper) at microwave frequencies, which have similar material properties as the PEC. The conclusions also apply to metals at higher frequencies, such as Drude metals with dispersion and loss, as long as the electric near field is dominated by the component perpendicular to the metal surface. To verify this point, we consider a silver sphere of radius $R_s$ under the excitation of an elliptical polarized plane wave same as in Figure 1a. The permittivity of silver is described by the Drude model $\varepsilon_{Ag} = 1 - \omega_p^2/(\omega^2 + i\omega\gamma)$ with $\omega_p = 1.393 \times 10^{16}$ rad/s and $\gamma = 3.145 \times 10^{13}$ rad/s[40]. We simulate the system at different wavelengths $\lambda = 15$ μm, 7 μm, and 3 μm while fixing the ratio $\lambda/R_s = 20$. Figure 6a shows the electric dark spots and C lines at the three wavelengths, respectively. Figures 6b, 6c, and 6d show the $|\mathbf{E}|^2/|\mathbf{E}_0|^2$, arg(ψ), and $|\mathbf{k}|/|\mathbf{k}_0|$ near the dark spot at the north pole. We notice that, at $\lambda = 15$ μm, the dark spots are nearly ideal topological dark spots characterized by a negligible electric field, a phase singularity of charge $q = 2$, and a very large local wavenumber. As the wavelength reduces, the electric field at the dark spots increases (Figure 6b), the phase singularity of charge $q = 2$ bifurcates into two phase singularities of charge $q = 1$ (Figure 6c), and the local wavenumber also reduces (Figure 6d). At the wavelength $\lambda = 3$ μm, the dark spots can be considered quasi-dark spots. The phenomena are attributed to the dispersion of the permittivity of the Drude silver. At the smaller wavelengths (i.e., $\lambda = 7$ μm and $\lambda = 3$ μm), the permittivity of silver has a small absolute value. As a result, the electric near field on the sphere's surface cannot be regarded as a scalar field due to the presence of the tangential components of the field. We note that the quasi-dark spots still have a very small electric field compared with the incident field with $|\mathbf{E}|^2/|\mathbf{E}_0|^2 < 1 \times 10^{-2}$. Thus, they are good approximations of the ideal dark spots for practical applications.

## 3. CONCLUSION

In conclusion, we demonstrate that metal scatterers can be employed to generate dark spots of electric near field. These dark spots are topologically protected and robust against small perturbations, and their positions can be controlled by tuning the direction and polarization of the incident light. We reveal that the topologically protected dark spots are accompanied by topological near-field vortices and superoscillation phenomena. The results provide a robust mechanism to generate and manipulate electric dark spots, which can find broad applications in super-resolution imaging, optical sensing, optical trapping, and optical metrology. Meta-optics has provided rich platforms for studying the topological properties of light fields in both real and momentum spaces[24,30,41]. Our results further enrich the physics of topological photonics in real space. Besides, the mechanism may be extended to other optical systems with extreme boundary conditions, such as the near-zero refractive index materials[42].

**Funding.** The work described in this paper was supported by grants from the National Natural Science Foundation of China (No. 12322416) and Research Grants Council of the Hong Kong Special Administrative Region, China (Project No. AoE/P-502/20).

**Acknowledgements.** We thank Dr. Ruo-Yang Zhang for the helpful discussions.

**Disclosures.** The authors declare no competing financial interest.

**REFERENCES**
(1) Vernon, A. J.; Dennis, M. R.; Rodríguez-Fortuño, F. J. 3D zeros in electromagnetic fields. *Optica* **2023**, *10* (9), 1231-1240.
(2) Allen, L.; Beijersbergen, M. W.; Spreeuw, R.; Woerdman, J. Orbital angular momentum of light and the transformation of Laguerre-Gaussian laser modes. *Phys. Rev. A* **1992**, *45* (11), 8185.
(3) Aharonov, Y.; Albert, D. Z.; Vaidman, L. How the result of a measurement of a component of the spin of a spin-1/2 particle can turn out to be 100. *Phys. Rev. Lett.* **1988**, *60* (14), 1351.
(4) Berry, M. V. Evanescent and real waves in quantum billiards and Gaussian beams. *J. Phys. A Math. Gen.* **1994**, *27* (11), L391.
(5) Berry, M.; Popescu, S. Evolution of quantum superoscillations and optical superresolution without evanescent waves. *J. Phys. A Math. Gen.* **2006**, *39* (22), 6965.
(6) Dennis, M. R.; Hamilton, A. C.; Courtial, J. Superoscillation in speckle patterns. *Opt. Lett.* **2008**, *33* (24), 2976-2978.
(7) Zheludev, N. I.; Yuan, G. Optical superoscillation technologies beyond the diffraction limit. *Nat. Rev. Phys.* **2022**, *4* (1), 16-32.
(8) Song, A. Y.; Catrysse, P. B.; Fan, S. Broadband control of topological nodes in electromagnetic fields. *Phys. Rev. Lett.* **2018**, *120* (19), 193903.
(9) Hell, S. W.; Wichmann, J. Breaking the diffraction resolution limit by stimulated emission: stimulated-emission-depletion fluorescence microscopy. *Opt. Lett.* **1994**, *19* (11), 780-782.
(10) Balzarotti, F.; Eilers, Y.; Gwosch, K. C.; Gynnå, A. H.; Westphal, V.; Stefani, F. D.; Elf, J.; Hell, S. W. Nanometer resolution imaging and tracking of fluorescent molecules with minimal photon fluxes. *Science* **2017**, *355* (6325), 606-612.
(11) Yuan, G. H.; Zheludev, N. I. Detecting nanometric displacements with optical ruler metrology. *Science* **2019**, *364* (6442), 771-775.
(12) Vetsch, E.; Reitz, D.; Sagué, G.; Schmidt, R.; Dawkins, S.; Rauschenbeutel, A. Optical interface created by laser-cooled atoms trapped in the evanescent field surrounding an optical nanofiber. *Phys. Rev. Lett.* **2010**, *104* (20), 203603.
(13) Zito, G.; Rusciano, G.; Sasso, A. Dark spots along slowly scaling chains of plasmonic nanoparticles. *Opt. Express* **2016**, *24* (12), 13584-13589.
(14) Palombo Blascetta, N.; Lombardi, P.; Toninelli, C.; van Hulst, N. F. Cold and hot spots: from inhibition to enhancement by nanoscale phase tuning of optical nanoantennas. *Nano Lett.* **2020**, *20* (9), 6756-6762.
(15) Xia, J.; Tang, J.; Bao, F.; Sun, Y.; Fang, M.; Cao, G.; Evans, J.; He, S. Turning a hot spot into a cold spot: polarization-controlled Fano-shaped local-field responses probed by a quantum dot. *Light Sci. Appl.* **2020**, *9* (1), 166.
(16) Lim, S. W. D.; Park, J.-S.; Meretska, M. L.; Dorrah, A. H.; Capasso, F. Engineering phase and polarization singularity sheets. *Nat. Commun.* **2021**, *12* (1), 4190.

(17) Vernon, A. J.; Rodríguez-Fortuño, F. J. Creating and moving nanoantenna cold spots anywhere. *Light Sci. Appl.* **2022**, *11* (1), 258.
(18) Lim, S. W. D.; Park, J.-S.; Kazakov, D.; Spägele, C. M.; Dorrah, A. H.; Meretska, M. L.; Capasso, F. Point singularity array with metasurfaces. *Nat. Commun.* **2023**, *14* (1), 3237.
(19) Dennis, M. R.; O'holleran, K.; Padgett, M. J. Singular optics: optical vortices and polarization singularities. In *Progress in optics*, Vol. 53; Elsevier, 2009; pp 293-363.
(20) Nye, J. F. Lines of circular polarization in electromagnetic wave fields. *Proc. R. Soc. A: Math. Phys. Eng. Sci.* **1983**, *389* (1797), 279-290.
(21) Peng, J.; Liu, W.; Wang, S. Polarization singularities in light scattering by small particles. *Phys. Rev. A* **2021**, *103* (2), 023520.
(22) Chen, W.; Yang, Q.; Chen, Y.; Liu, W. Evolution and global charge conservation for polarization singularities emerging from non-Hermitian degeneracies. *PNAS* **2021**, *118* (12), e2019578118.
(23) Jia, S.; Peng, J.; Cheng, Y.; Wang, S. Chiral discrimination by polarization singularities of a metal sphere. *Phys. Rev. A* **2022**, *105* (3), 033513.
(24) Peng, J.; Zhang, R.-Y.; Jia, S.; Liu, W.; Wang, S. Topological near fields generated by topological structures. *Sci. Adv.* **2022**, *8* (41), eabq0910.
(25) Zhou, M.; You, S.; Liu, J.; Qin, K.; Wang, J.; Zhang, Y.; Xiang, H.; Zhou, C.; Han, D. Selective Perturbation of Eigenfield Enables High-Q Quasi-Bound States in the Continuum in Dielectric Metasurfaces. *ACS Photonics* **2024**.
(26) Dennis, M. R. Rows of optical vortices from elliptically perturbing a high-order beam. *Opt. Lett.* **2006**, *31* (9), 1325-1327.
(27) Forbes, A.; de Oliveira, M.; Dennis, M. R. Structured light. *Nat. Photonics* **2021**, *15* (4), 253-262.
(28) Shen, Y.; Wang, X.; Xie, Z.; Min, C.; Fu, X.; Liu, Q.; Gong, M.; Yuan, X. Optical vortices 30 years on: OAM manipulation from topological charge to multiple singularities. *Light Sci. Appl.* **2019**, *8* (1), 90.
(29) Spaegele, C. M.; Tamagnone, M.; Lim, S. W. D.; Ossiander, M.; Meretska, M. L.; Capasso, F. Topologically protected optical polarization singularities in four-dimensional space. *Sci. Adv.* **2023**, *9* (24), eadh0369.
(30) Yang, B.; Guo, Q.; Wang, D.; Wang, H.; Xia, L.; Xu, W.; Kang, M.; Zhang, R.-Y.; Zhang, Z.-Q.; Zhu, Z.; et al. Scalar topological photonic nested meta-crystals and skyrmion surface states in the light cone continuum. *Nat. Mater.* **2023**, *22* (10), 1203-1209.
(31) Dennis, M. R. Topological singularities in wave fields. University of Bristol, 2001.
(32) Guo, C.; Li, J.; Xiao, M.; Fan, S. Singular topology of scattering matrices. *Phys. Rev. B* **2023**, *108* (15), 155418.
(33) Berry, M.; Dennis, M. Phase singularities in isotropic random waves. *Proc. R. Soc. A: Math. Phys. Eng. Sci.* **2000**, *456* (2001), 2059-2079.
(34) Berry, M.; Zheludev, N.; Aharonov, Y.; Colombo, F.; Sabadini, I.; Struppa, D. C.; Tollaksen, J.; Rogers, E. T.; Qin, F.; Hong, M. Roadmap on superoscillations. *J. Opt.* **2019**, *21* (5), 053002.
(35) Forbes, K. A.; Andrews, D. L. Optical orbital angular momentum: twisted light and chirality. *Opt. Lett.* **2018**, *43* (3), 435-438.
(36) Brullot, W.; Vanbel, M. K.; Swusten, T.; Verbiest, T. Resolving enantiomers using the optical angular momentum of twisted light. *Sci. Adv.* **2016**, *2* (3), e1501349.
(37) Nye, J. F.; Berry, M. V. Dislocations in wave trains. *Proc. R. Soc. A: Math. Phys. Eng. Sci.* **1974**, *336* (1605), 165-190.
(38) Xu, H.; Bjerneld, E. J.; Käll, M.; Börjesson, L. Spectroscopy of single hemoglobin molecules by surface enhanced Raman scattering. *Phys. Rev. Lett.* **1999**, *83* (21), 4357.
(39) Yang, Q.; Chen, W.; Chen, Y.; Liu, W. Symmetry protected invariant scattering properties for incident plane waves of arbitrary polarizations. *Laser Photonics Rev.* **2021**, *15* (6), 2000496.
(40) Ordal, M. A.; Bell, R. J.; Alexander, R. W.; Long, L. L.; Querry, M. R. Optical properties of fourteen metals in the infrared and far infrared: Al, Co, Cu, Au, Fe, Pb, Mo, Ni, Pd, Pt, Ag, Ti, V, and W. *Appl. Opt.* **1985**, *24* (24), 4493-4499.
(41) Fu, T.; Zhang, R.-Y.; Jia, S.; Chan, C. T.; Wang, S. Near-Field Spin Chern Number Quantized by Real-Space Topology of Optical Structures. *Phys. Rev. Lett.* **2024**, *132* (23), 233801.
(42) Liberal, I.; Engheta, N. Near-zero refractive index photonics. *Nat. Photonics* **2017**, *11* (3), 149-158.

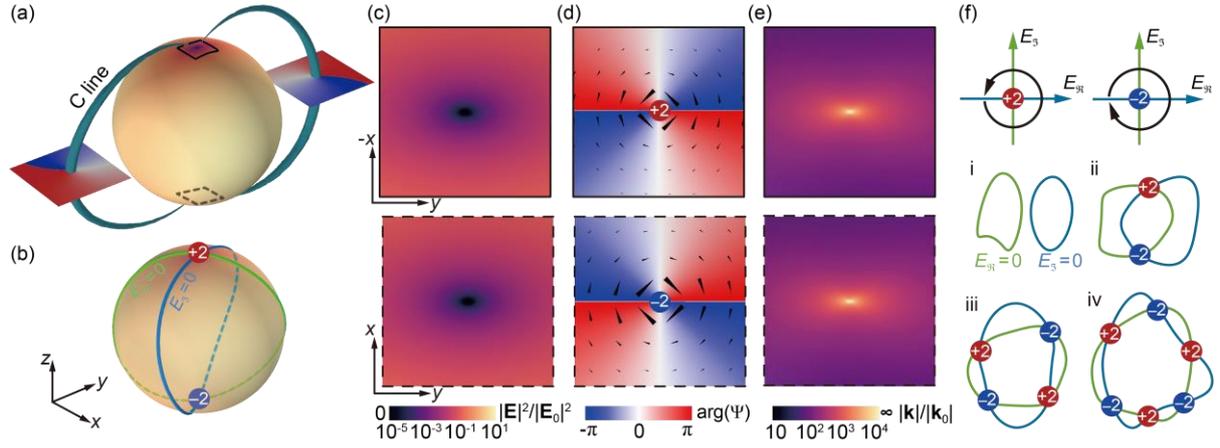

**Figure 1. Electric dark spots on the surface of a PEC sphere**. (a) Dark spots and C lines of electric fields. (b) The topological charges of dark spots and the lines of $E_\Re = 0$ and $E_\Im = 0$. (c) The normalized electric field intensity, (d) the phase $\arg(\Psi)$ and local wave vectors **k** (black arrows), and (e) the normalized wavenumber in the area enclosed by the black boxes in (a). The upper (lower) panels correspond to the solid line (dashed line) box. (f) The topological charges in the complex plane and four example types of the dark spots at the intersections of $E_\Re = 0$ and $E_\Im = 0$. The sphere has a radius of 100 nm. The frequency is $f = 150$ THz. The incident light is $\mathbf{E}_0 = (\hat{\mathbf{x}} + i0.6\hat{\mathbf{y}})e^{ikz}$.

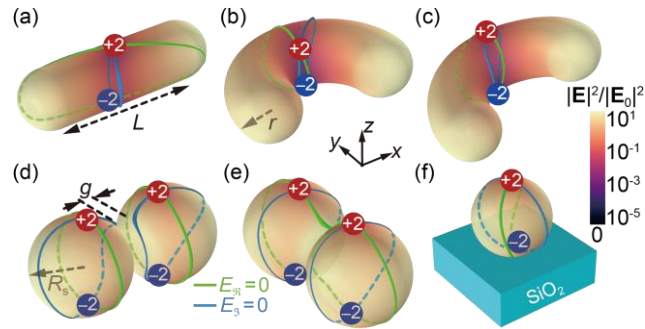

**Figure 2. The robustness of the dark spots upon perturbations**: The lines of $E_\Re = 0$ and $E_\Im = 0$ on (a) nanorod, (b) bent nanorod, (c) twisted nanorod, (d-e) coupled spheres, and (f) sphere on a SiO$_2$ substrate of infinite thickness. In (a-c), we set $r = 100$ nm and the nanorod length $L = 500$ nm. The pitch of the twisted nanorod in (c) is 300 nm. The gap between the two spheres in (d) and (e) is $g = 5$ nm. The relative permittivity of SiO$_2$ is $\varepsilon_r = 2.25$. The incident wave is the same as that in Figure 1.

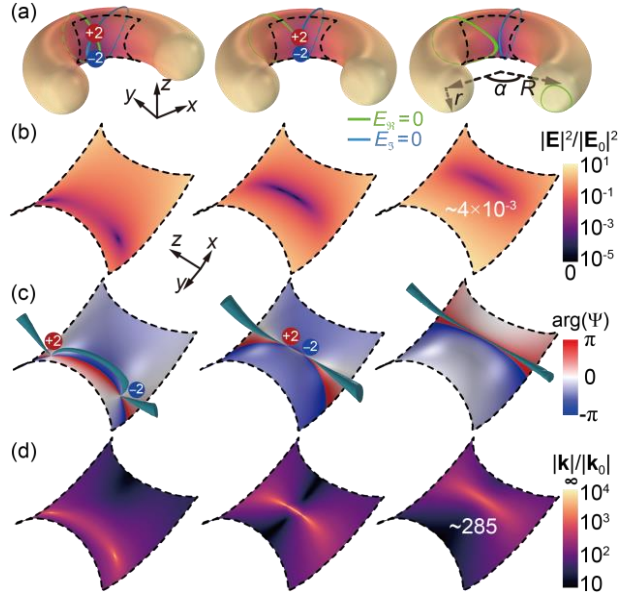

**Figure 3. Topological transition of the dark spots.** (a) The dark spots on the different split-ring geometries with $\alpha = 0.9\pi$, $0.66\pi$, and $0.6\pi$. (b) The electric field distributions, (c) the phase $\arg(\Psi)$ and C lines, and (d) the local wave number in the region enclosed by the black dashed box in (a). The split ring geometries have identical inner and outer radii $r = 70$ nm and $R = 175$ nm. The incident wave is identical to that in Figure 1.

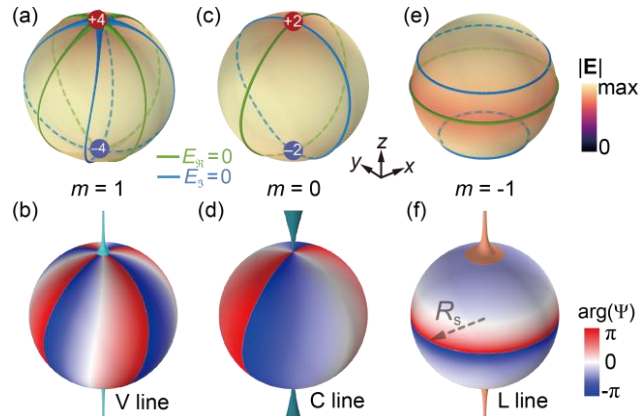

**Figure 4. Higher-order dark spots.** Dark spots and the lines of $E_\Re = 0$ and $E_\Im = 0$ on the PEC sphere excited by the right-handed circularly polarized Bessel beams of (a) $m = 1$, (c) $m = 0$, and (e) $m = -1$. (b) The V line and phase $\arg(\Psi)$ corresponding to (a). (d) The C line and phase $\arg(\Psi)$ corresponding to (c). (f) The L line and phase $\arg(\Psi)$ corresponding to (e). The radii of the spheres are $R_s = 100$ nm.

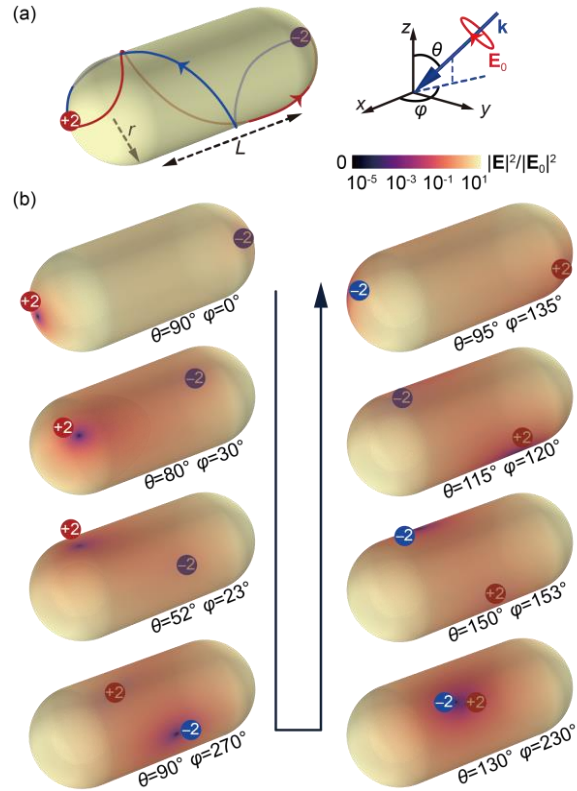

**Figure 5. Manipulating the position of the dark spots.** (a) The trajectories of the dark spots on the nanorod surface. The incident directions are defined in the inset of (a). (b) Simulated results for dark spots on the nanorod surface for different polar ($\theta$) and azimuthal ($\varphi$) angles of the incident wave. The incident wave is circularly polarized. The nanorod has length $L = 300$ nm and radius $r = 100$ nm.

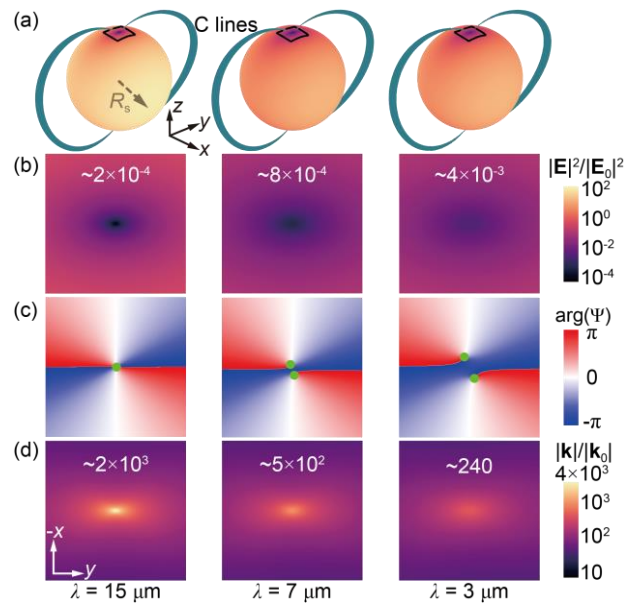

**Figure 6. The electric dark spots on silver spheres.** (a) The C lines and dark spots on the silver sphere surface at different wavelengths $\lambda = 15$ μm, $\lambda = 7$ μm, and $\lambda = 3$ μm (from left to right). The (b) electric field intensity, (c) dynamical phase, and (d) wave number in the region enclosed by the black box in (a) near the dark spot. The radii of the silver spheres are $R_s = 750$ nm, $R_s = 350$ nm, and $R_s = 150$ nm from left to right. The incident light is identical to that in Figure 1.